\documentclass[twocolumn,noshowpacs,preprintnumbers,amsmath,amssymb]{revtex4}

\usepackage{graphicx}
\usepackage{dcolumn}
\usepackage{bm}
\usepackage{amsfonts, color, soul}

\newcommand{\phn}{phosphorylation}
\newcommand{\phd}{phosphorylated}
\newcommand{\be}{\begin{equation}}
\newcommand{\ee}{\end{equation}}
\newcommand{\bea}{\begin{eqnarray}}
\newcommand{\eea}{\end{eqnarray}}

\newcommand{\cact}[1]{[{\rm C}_{#1}]}

\newcommand{\acon}{[{\rm A}]}
\newcommand{\at}{\acon_{\rm T}}
\newcommand{\ct}{ {\rm [C]_T} }

\begin{document}

\title{An allosteric model of circadian KaiC phosphorylation}

\date{\today}

\author{Jeroen S. van Zon}
\affiliation{Department of Mathematics, Imperial College London,
London SW7 2AZ, UK} 
\affiliation{Division of Physics and Astronomy, Vrije Universiteit Amsterdam,
1081 HV Amsterdam, The Netherlands}

\author{David K. Lubensky}
\altaffiliation[Present address: ]{Department of Physics, University of
Michigan, Ann Arbor MI 48109-1040}
\affiliation{Division of Physics and Astronomy, Vrije Universiteit Amsterdam,
1081 HV Amsterdam, The Netherlands}

\author{Pim R. H. Altena}
\affiliation{FOM Institute for Atomic and Molecular Physics,
Kruislaan 407, 1098 SJ Amsterdam, The Netherlands}

\author{Pieter Rein ten Wolde}
\email[email address: ]{tenwolde@amolf.nl}
\affiliation{FOM Institute for Atomic and Molecular Physics,
Kruislaan 407, 1098 SJ Amsterdam, The Netherlands}

\begin{abstract}
In a recent series of ground-breaking experiments, Nakajima {\em et
  al.} [{\em Science} {\bf 308}, 414-415 (2005)] showed that the three
cyanobacterial clock proteins KaiA, KaiB, and KaiC are sufficient {\em
  in vitro} to generate circadian phosphorylation of KaiC.  Here, we
present a mathematical model of the Kai system.  At its heart is the
assumption that KaiC can exist in two conformational states, one
favoring phosphorylation and the other dephosphorylation.  Each
individual KaiC hexamer then has a propensity to be phosphorylated in
a cyclic manner. To generate macroscopic oscillations, however, the
phosphorylation cycles of the different hexamers must be
synchronized. We propose a novel synchronisation mechanism based on
{\em differential affinity}: KaiA stimulates KaiC phosphorylation, but
the limited supply of KaiA dimers binds preferentially to those KaiC
hexamers that are falling behind in the oscillation.  KaiB sequesters
KaiA and stabilizes the dephosphorylating KaiC state.  We show that
our model can reproduce a wide range of published data, including {the
observed insensitivity of the oscillation period to variations in
temperature}, and that it makes nontrivial predictions about the
effects of varying the concentrations of the Kai proteins.
\end{abstract}

\maketitle

Cyanobacteria are the simplest organisms to use circadian rhythms to
anticipate the changes between day and night. In the cyanobacterium
{\em Synechococcus elongatus}, the three genes {\em kaiA},
{\em kaiB} and {\em kaiC} are the central components of the circadian
clock \cite{ishiura98}. In higher organisms, it is believed that circadian
rhythms are driven primarily by transcriptional
feedback\cite{dunlap04}. KaiC phosphorylation, however, shows a
circadian rhythm even when transcription and translation
are inhibited \cite{tomita05}. Still more remarkably, it was recently shown that this rhythmic
KaiC phosphorylation can be reconstituted {\em in vitro} in the
presence of only KaiA, KaiB, and ATP \cite{nakajima05}.  The Kai system
thus represents a very rare example of a functional biochemical
circuit that can be re-created in the test tube.  It is a major
open question to explain how stable oscillations can result from the
experimentally observed interactions among the different Kai
proteins.

In living cells, KaiC phosphorylation increases during the subjective
day and decreases during the subjective night, and this
phosphorylation in turn regulates KaiC's activity as a global
transcriptional repressor~\cite{nakahira04}.  KaiC forms a hexamer
both {\em in vivo} and {\em in vitro} \cite{kageyama03}; KaiA is
present in the cell as a dimer \cite{kageyama03} and KaiB as a dimer
\cite{kageyama03, kageyama06} or a tetramer \cite{hitomi05}.  KaiC has
both auto-dephosphorylation and weaker auto-phosphorylation activity,
with the latter dependent on ATP
binding~\cite{hayashiitoh04,nishiwaki00,iwasaki02,kitayama03,xu03,nishiwaka04}.
KaiC phosphorylation is stimulated by
KaiA~\cite{iwasaki02,xu03,hayashi04}, whereas KaiB appears to interfere
with this effect~\cite{iwasaki02,williams02,kitayama03,xu03}.  KaiC hexamers
form heteromultimeric complexes with KaiA and KaiB dimers, but one
such complex contains no more than one KaiC
hexamer~\cite{iwasaki99,kageyama03,kageyama06}.  The composition of
these complexes varies with a roughly 24 hour period.

Nakajima {\em et al.}'s striking observation of {\em in vitro}
oscillations~\cite{nakajima05} in KaiC \phn\ poses an obvious
challenge for modelers.  Not only is there the potential for detailed
comparisons between a model's predictions and the wealth of
experimental data, the Kai system also has several novel features.
{Most notably, ATP is consumed, and the system is driven out of equilibrium, only through the repeated phosphorylation and dephosphorylation of KaiC.  Other reactions, such as the (un)binding of KaiA and KaiB to KaiC, should thus obey detailed balance.}
Moreover, {unlike} in most biological
oscillations~\cite{goldbeter96}{, in the Kai system the proteins are neither
created nor destroyed.  This
imposes} significant constraints on any model that hopes to explain the
{\em in vitro} oscillations.

Several previous papers have put forward interesting ideas on how
these oscillations might occur~\cite{emberly06,mehra06,kurosawa06}.
{However, they either require that KaiC hexamers can bind to each
  other to form higher order complexes}~\cite{emberly06,mehra06}, {
  a
possibility ruled out by recent experiments}
\cite{kageyama03,kageyama06} {, or they assume that KaiA and KaiB
  can each take on multiple forms}~\cite{kurosawa06}.  In the latter
case, the authors propose that these forms may correspond to different
subcellular localizations, but that suggestion cannot hold for the
{\em in vitro} system. 
Emberly and Wingreen~\cite{emberly06} introduced the
elegant hypothesis that exchange of monomers among KaiC hexamers might
contribute to oscillations, an idea supported by recent
observations~\cite{kageyama06}.  Their own work, however, shows that
such exchange by itself is insufficient to produce sustained
oscillations.  There is thus clearly another mechanism at work in the
Kai system.

Here, we propose such a mechanism.  Our model is built upon two key
elements.  First, we hypothesize that an isolated KaiC hexamer already
has a tendency to be cyclically phosphorylated and dephosphorylated as
it flips between two allosteric states.  Second, we suggest that these
noisy oscillations of individual hexamers can be synchronized through
the phenomenon of {\em differential affinity}, whereby the laggards in
a population outcompete the other hexamers for a limited number of
KaiA molecules that stimulate phosphorylation.  The slowest hexamers
thus speed up while the fastest are forced to slow down, causing the
entire population to oscillate in phase. 

In the rest of this paper, we first, in section I, show how a simple
picture of allosteric transitions in KaiC leads each hexamer to have
an intrinsic phosphorylation cycle.  We then use an idealized model to
introduce the concept of differential affinity in section II. {This
  model shows that the mechanism requires only a few
  generic ingredients, suggesting that the same synchronization
  principle could be at work in other biological systems. Finally, we
  turn in section III to a more complicated model of the Kai
  system. This model reproduces the phosphorylation behavior of KaiC
  not only in the} {\em in vitro} {experiments in which all three
  Kai proteins are present, but also in systems where KaiA and/or KaiB
  are absent. In fact, we found that the experiments on
  the various subsets of the three Kai proteins strongly constrain the model's design. Beyond synchronizing oscillations, KaiA and KaiB must also bind to and stabilize one or the other KaiC state.}  When this binding is
strong enough, the system moreover exhibits temperature compensation,
as observed~\cite{nakajima05}.

\section{Allosteric Model}

In this section, we introduce a simple model of allosteric transitions
in KaiC that naturally gives rise to repeated rounds of
phosphorylation and dephosphorylation within each hexamer.  Allosteric
conformational changes are widespread in biochemistry, and the
conformations of members of the RecA/DnaB superfamily, to which KaiC
belongs, have been extensively studied~\cite{wang04,
bell05}. 

{\bf KaiC monomers} Although there is strong evidence that KaiC {\em
monomers} can be \phd\ at multiple sites~\cite{xu04, nishiwaki04,
iwasaki02}, most published data do not distinguish between different
\phd\ forms.  We thus assume that KaiC monomers have only two
\phn\ states, \phd\ and un\phd.

\begin{figure}
\centering
\includegraphics[width=\columnwidth]{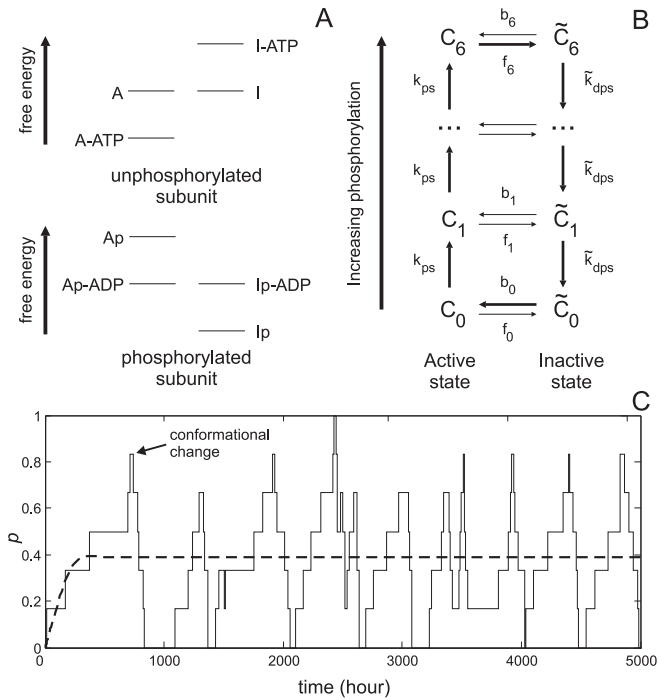}
\caption{
(A) Schematic free energy levels for KaiC
subunits.  Subunits can be in the active (A) or the inactive (I)
state. Furthermore, subunits can be phosphorylated (p) and bind ATP or
ADP. Phosphorylation favors the inactive state, nucleotide
binding the active state.  (B) Reaction network for a KaiC hexamer
with 6 phosphorylation sites. ${\rm C}_i$ and $\widetilde{\rm C}_i$
denote a hexamer with
$i$ phosphorylated monomers in respectively the active and inactive state.  (C) Phosphorylation cycles for the
model in B. The phosphorylation level $p$ of a single KaiC
hexamer, as obtained by stochastic simulations (solid line), and of a
population of hexamers, obtained from the mean-field rate equations
(dashed line). The phosphorylation level $p \equiv \sum_i i([{\rm
C}_i]+[\widetilde{\rm C}_i])/\sum_i 6 ([{\rm C}_i]+[\widetilde{\rm
C}_i])$, where $[{\rm C}_i]$ is the concentration of hexamers in state
${\rm C}_i$. {For parameter values, see} {\em Supporting
Information}.\label{fig:fig1}}
\end{figure}

We postulate that an individual KaiC monomer can be in either an
active (A) or an inactive (I) conformation. Fig.~\ref{fig:fig1}A shows
the free energies of the different monomer states; we consider ATP
binding only to un\phd\ and ADP binding only to \phd\ monomers. As the
figure indicates, we assume that \phn\ favors the {\em inactive} over the active state. Nucleotides have a higher affinity for
monomers in the active state than for those in the inactive state, so
nucleotide binding favors the {\em active} state over the
inactive one.  We also take both the transfer of a phosphate from ATP
to a KaiC monomer and the removal of the phosphate from the monomer to
be thermodynamically favorable.  Taken together, these elements allow
for a phosphorylation cycle: {unphosphorylated monomers prefer to be in
the A state, where ATP hydrolysis drives phosphorylation, while
phosphorylated monomers prefer to be in the I state, where
dephosphorylation occurs spontaneously}. Each monomer thus tends to go
through the sequence of reactions A $\rightarrow$ A-ATP $\rightarrow$
Ap-ADP $\rightarrow$ Ip-ADP $\rightarrow$ Ip $\rightarrow$ I
$\rightarrow$ A, during which one ATP molecule is hydrolyzed.

{\bf KaiC hexamers}.  In the spirit of the 
MWC model~\cite{monod65}, we assume that the
energetic cost of having two different monomer conformations in the
same hexamer is prohibitively large.  We can then speak of a hexamer
as being in either the A or the I state.  The total (free) energy of
the hexamer is simply the sum of the contributions from its
constituent monomers.  Highly \phd\ hexamers thus prefer to be in the
I state, where they will be dephosphorylated, while weakly \phd\
hexamers prefer the A state, where they will be phosporylated.  As a
result, each hexamer tends to go through a cycle in which it is first
phosphorylated, then dephosphorylated, as indicated in
Fig.~\ref{fig:fig1}B and Fig.2 of the {\em Supporting Information}.

The transition (or {\em flip}) rates $f_i$ for a hexamer with $i$
\phd\ monomers to go from the A to the I state and $b_i$ to go from the I
to the A state depend on the energy barriers to the conformational changes.  If we assume that ATP and
ADP exchange are fast, so that the free energy of each state is
well-defined, then the difference in free energy $\Delta G$ between
the I and A states grows linearly with $i$:  $\Delta G(i) =
i \Delta G_{\rm p} + (6-i) \Delta G_{\rm u}$, where the subscripts p
and u refer to the free-energy differences for phosphorylated and
unphosphorylated monomers, respectively.  The natural phenomenological
assumption is then that the flip rates depend exponentially on the
free-energy difference:
\begin{alignat}{3}
f_i & =  k_0 \exp[\Delta G(i)/2]  &\,\sim \,& c^i  \label{eq:fwdrate} \\
b_i & =  k_0 \exp[-\Delta G(i)/2] &\, \sim \,&  c^{-i} \; , \label{eq:bkrate}
\end{alignat}
where $k_0$ sets the basic timescale and $ c= \exp[(\Delta G_{\rm p} -
\Delta G_{\rm u})/2]$.  Alternatively, one can develop an explicit
transition state theory that includes the number of bound nucleotides
as one of the order parameters for the conformational transition (see
{\em Supporting Information}).  This leads to flipping rates that vary
exponentially with $i$ just as in
Eqs.~\ref{eq:fwdrate}--\ref{eq:bkrate}.  In either case, the rates
depend strongly on the phosphorylation level, with the consequence
that hexamers can flip from A to I only when most of their monomers
are \phd\ and from I to A only when most are not \phd.

In Fig.~\ref{fig:fig1}C, we show the time dependence of the
phosphorylation level of a single KaiC hexamer obtained by {
  Monte Carlo simulations of the chemical master equation (see} {\em
  Supporting Information})\cite{gillespie77}.  Initially, the KaiC hexamer is
in the unphosphorylated active state, $C_0$. KaiC (de)phosphorylation
clearly occurs in a cyclic fashion, with few transitions from one
conformation to the other occurring at intermediate
phosphorylation. However, both the amplitude and the period of the
phosphorylation cycle are highly variable. Due to this variability,
the phosphorylation cycles of a population of independent KaiC hexamers will
quickly dephase. As a result, in
Fig.~\ref{fig:fig1}C the mean phosphorylation level of the KaiC
population calculated {by integrating deterministic rate
equations based on the law of mass action}
shows no oscillatory behavior. In order to explain the oscillations
observed in the {\em in vitro} Kai system, the uncoupled
phosphorylation
cycles of the individual KaiC hexamers need to be synchronized.\\

\section{Synchronisation with differential affinity}
\label{sec:II}

The natural candidates to link the phosporylation states of different
KaiC hexamers are the other two Kai proteins.  Here, we present a
simple model in which KaiA plays this role by catalyzing
phosphorylation in the active state, while KaiB is completely absent.
This model will allow us to introduce several important ideas without
the distractions that a more faithful description would entail.  It
shows synchronized limit-cycle oscillations in KaiC phosphorylation,
provided that the concentration of KaiA is sufficiently small and that
KaiA binds to KaiC with {\em differential affinity}: KaiA should bind
most strongly to weakly phosphorylated KaiC hexamers.  Although here
we limit our discussion to a particular model inspired by the Kai
system, the differential affinity mechanism is also amenable to a more
general, abstract formulation, described in the {\em
  Supporting Information}, which also shows that the oscillations
arise through a supercritical Hopf bifurcation.

We assume that only a single dimer of KaiA can bind to a KaiC
hexamer, and we force every hexamer to proceed through the states
${\rm C}_0$--${\rm C}_6$ and $\widetilde{{\rm C}}_6$--$\widetilde{{\rm
    C}}_0$ in order (thus neglecting intermediate flips).
This yields
\begin{alignat}{1}
& {\rm C}_6 \overset{f_6}{\rightarrow}
  \widetilde{\rm C}_6, \,\,\,\,\,\, \widetilde{\rm C}_0 \overset{b_0}{\rightarrow} {\rm C}_0
  \label{eq:S1}\\
& \widetilde{{\rm C}}_i \overset{\tilde{k}_{{\rm dps}}}{\rightarrow}
\widetilde{\rm C}_{i-1}\label{eq:S2} \\
&{\rm C}_i + {\rm A} \underset{k_i^{\rm Ab}}{\overset{k^{\rm Af}}{\rightleftarrows}} {\rm AC}_i  
\overset{k_{\rm pf}}{\rightarrow} {\rm C}_{i+1} + {\rm A}  \,\,\,\,\, (i \neq 6) \; .
 \label{eq:S3}
\end{alignat}
\noindent We use
deterministic, mass-action kinetics to model the effects of
these reactions.  Here ${\rm C}_i$ and $\widetilde{\rm C}_i$ denote $i$-fold
phosphorylated KaiC hexamers in the active and inactive states, and A
denotes a KaiA dimer.  Eqs.~\ref{eq:S1}-\ref{eq:S3} describe the same processes within a
single hexamer as the diagram in Fig.~\ref{fig:fig1}B, with the exception that \phn\ of the active state now requires KaiA, which associates with active KaiC with on and off rates $k^{{\rm Af}}$ and
$k_i^{{\rm Ab}}$ and stimulates phosphorylation
with a rate $k_{\rm pf}$ (Eq.~\ref{eq:S3}).
We implement differential affinity by setting $k_i^{{\rm Ab}} =
k_0^{\rm Ab} \alpha^{i}$, with
$\alpha > 1$ (see {\em Supporting Information}).

\begin{figure}
\centering
\includegraphics[width=\columnwidth]{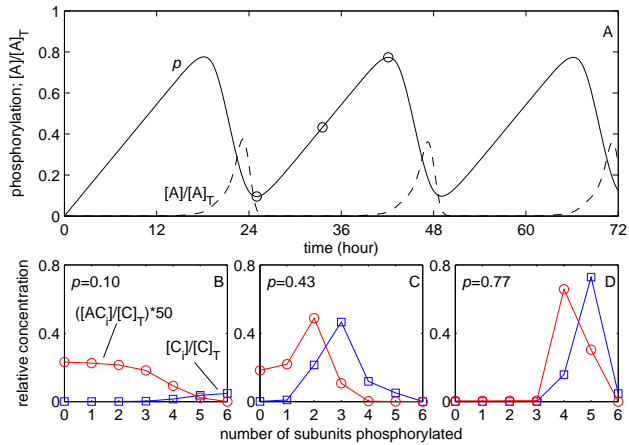}
\caption{
Limit cycle oscillations in KaiC phosphorylation for the
    simplified model defined by Eqs.~\ref{eq:S1}-\ref{eq:S3}
  (A). Mean phosphorylation level $p$ and {normalized concentration
    of free KaiA} $\acon/\at$.  During the phosphorylation phase [A]
  drops almost to zero. (B), (C) and (D) KaiA binding at 3 stages of
  the phosphorylation phase, marked by circles in (A). KaiA favors the
  less phosphorylated KaiC hexamers.  ${\rm [C]_T}$, total KaiC
  concentration; ${\rm AC}_i$, {complex of KaiA and} ${\rm C}_i$.
  We take $\at/\ct = 0.02$ {and initially set} $\cact{0} = \ct$;
  see {\em Supporting Information} {for other
    parameters}. \label{fig:fig2}
}
\end{figure}

Fig.~\ref{fig:fig2}A shows the mean phosphorylation level of a
population of KaiC hexamers as a function of time. In contrast to the
behavior seen in Fig.~\ref{fig:fig1}C, there are clear oscillations:  The KaiA dimers
effectively couple the phosphorylation cycles of the individual KaiC
hexamers.  During the phosphorylation phase of the oscillations, most hexamers are in the active form. In this state they can bind KaiA, which stimulates their phosphorylation. The
concentration of KaiA, however, is limited; indeed, in this part of
the cycle the concentration of {\em free} KaiA is close to zero
(Fig.~\ref{fig:fig2}A). This means that the KaiC hexamers compete with
one another for KaiA. In this competition, the complexes with a lower
degree of phosphorylation have the advantage, because they have a
higher affinity for KaiA. Hence, during the phosphorylation phase,
KaiA will be mostly bound to the lagging hexamers. This is shown in
Fig.~\ref{fig:fig2}B, where the concentrations $[{\rm C}_i]$ and
$[{\rm AC}_i]$ are plotted versus $i$ for three different time points.  The distributions do
not overlap: KaiA has a clear preference for the less phosphorylated
KaiC hexamers. Since the phosphorylation rate depends on the amount
of bound KaiA, laggards with a low degree of phosphorylation will be
phosphorylated at a high rate, whereas front runners with a high
degree of phosphorylation will be unable to increase their \phn\ level
further. This is the essence of the differential affinity synchronisation mechanism.  

\section{Full Model of the Kai system}
The simple model of the previous section showed how differential
affinity can synchronize the oscillations of the different KaiC
hexamers. This model, however, neglects KaiB completely and is not
consistent with the large body of experimental data on the Kai
system. Here, we present a more refined allosteric model.

\subsection{The model}
The key ingredients of our model are:

\noindent{\em 1. KaiA can bind to the active form of KaiC,
  stimulating KaiC \phn.} \\ {Recent experiments suggest that, in the
absence of KaiB, KaiA binds as a single dimer to the CII domain of the
KaiC hexamer} \cite{pattanayek06}{. Since this is the domain containing KaiC's phosphorylation site, it seems reasonable that the
affinity of KaiA might depend on the phosphorylation state of
KaiC. We thus assume, as before, that a single KaiA can bind to the active state
of KaiC and that the affinity of KaiA for active KaiC decreases as the
phosphorylation level increases.}

\noindent {\em 2. The active state of KaiC is more stable than the
  inactive one.}\\ The experiments of~\cite{tomita05,kageyama06} show
that in the presence of only KaiA, KaiC becomes very highly \phd. In
the absence of KaiB, KaiC should thus have no tendency to cyclically
phosphorylate and dephosphorylate. {This requires that the active
  state of KaiC has a lower free energy than the inactive one (thus
  shifting the energy levels in Fig.}  \ref{fig:fig1}A {from their
  symmetric
  values).}  \\
\noindent {\em 3. KaiB can bind to the inactive form of KaiC. The
  resulting KaiB-KaiC complex can then bind to and sequester KaiA.} \\ The phosphorylation behavior of
  KaiC in the presence of KaiB, but not KaiA, is essentially identical
  to that of KaiC in the absence of both KaiA and
  KaiB~\cite{xu03,kitayama03}. This observation strongly suggests that
  KaiB does not directly affect \phn\ and de\phn\ rates. We propose instead the following functions for KaiB: 1)
  KaiB can increase the stability of the inactive state of KaiC by
  binding to it. This restores the capacity of individual KaiC
  hexamers to sustain phosphorylation cycles.
2) Strong binding of KaiA by KaiB associated with
  the inactive KaiC hexamers reduces the concentration of free KaiA
  dimers. This leads to a variant of the differential affinity
  mechanism, which is necessary for synchronizing the oscillations of
  the different KaiC hexamers, as we clarify below. Based upon the
  measured size of the heteromultimeric
  complexes~\cite{kageyama03,kageyama06}, we assume that the
  inactive form of KaiC can bind two KaiB dimers, and that
{  ${\rm B}_2\widetilde{\rm C}_4, {\rm B}_2\widetilde{\rm C}_3, {\rm B}_2\widetilde{\rm C}_2$, and ${\rm B}_2\widetilde{{\rm C}}_1$} can each bind two KaiA
  dimers with high affinity. Neither assumption is critical: a
  model in which more than two KaiB and two KaiA dimers can bind also
  generates oscillations.

{\em \noindent 4. The rate of spontaneous phosphorylation is lower than that of
  spontaneous dephosphorylation.}\\ The model includes spontaneous
  phosphorylation and dephosphorylation of both active and
  inactive KaiC. Since KaiC reaches a low
  phosphorylation level in the absence of KaiA (and KaiB)~\cite{xu03,tomita05,kageyama06}, the rate of
  spontaneous phosphorylation is lower than that of spontaneous
  dephosphorylation.

This model is described by the following reactions:
\begin{alignat}{1}
&{\rm C}_i \underset{b_i}{\overset{f_i}{\rightleftarrows}}
  \widetilde{{\rm C}}_i \label{eq:F0}\\
&{\rm C}_i + {\rm A} \overset{K_i}{\rightleftarrows} {\rm AC}_i
  \overset{\tilde{k}_{\rm pf}}{\rightarrow} {\rm C}_{i+1} + {\rm A}
\label{eq:F1} \\
&\widetilde{\rm C}_i + 2 {\rm B} \underset{k^{\rm Bb}_i}{\overset{k^{\rm Bf}_i}{\rightleftarrows}}
{\rm B}_2\widetilde{\rm C}_i, \,\,\,{\rm B}_2\widetilde{\rm C}_i + 2
          {\rm A} \overset{\widetilde{K}_i}{\rightleftarrows}
{\rm A}_2{\rm B}_2\widetilde{\rm C}_{i} \label{eq:F2}\\
&C_i \underset{k_{\rm dps}}{\overset{k_{\rm ps}}{\rightleftarrows}} {\rm C}_{i+1},
\,\,\,
\widetilde{{\rm C}}_i
\underset{\tilde{k}_{\rm dps}}{\overset{\tilde{k}_{\rm ps}}{\rightleftarrows}}
\widetilde{{\rm C}}_{i+1}
\label{eq:F3} \\
&{\rm B}_2\widetilde{{\rm C}}_i
\underset{\tilde{k}_{\rm dps}}{\overset{\tilde{k}_{\rm ps}}{\rightleftarrows}}
{\rm B}_2\widetilde{{\rm C}}_{i+1},
\,\,\,
{\rm A}_2{\rm B_2}\widetilde{{\rm C}}_i
\underset{\tilde{k}_{\rm dps}}{\overset{\tilde{k}_{\rm ps}}{\rightleftarrows}}
{\rm A}_2{\rm B_2}\widetilde{{\rm C}}_{i+1} \; . \label{eq:F4}
\end{alignat}
{As in} Section~\ref{sec:II}, {we assume the reaction rates are
  given by deterministic, mass-action kinetics. The most critical
  parameters are the (de)phosphorylation rates. They have not
  been directly measured, but they are strongly constrained by
  the large number of quantitative} {\em in vitro}{ experiments on
  the subsets of Kai proteins
(see below).  The model's predictions are much less
  sensitive to the remainder of its 39 parameters; for these we have
  simply chosen plausible values} (see {\em Supporting
  Information}).

\begin{figure}[t]
\centering
\includegraphics[width=\columnwidth]{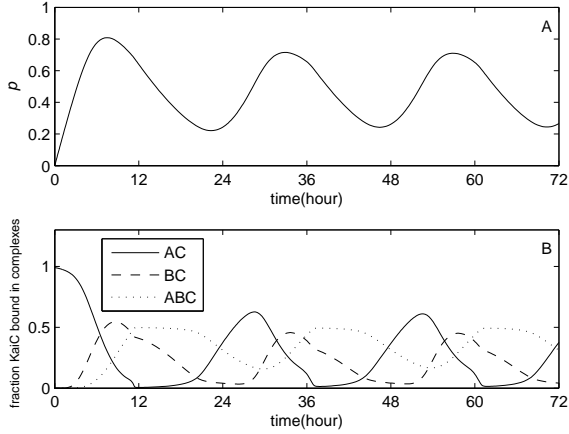}
\caption{Sustained oscillations in the full model defined by
  Eqs.~\ref{eq:F0}-\ref{eq:F4}.  (A) The mean phosphorylation level
  $p$ of KaiC shows a stable 24hr. rhythm. (B) Kai complexes. {At
    $t=0$, KaiC is fully unphosphorylated; ${\rm [C_0]
      = [C]_T; [A] = [A]_T; [B] = [B]_T}$. KaiA then binds KaiC and
    stimulates phosphorylation}. Next, the amount of KaiB-KaiC complex
  (${\rm [BC]}$) increases at high phosphorylation as KaiB binds to
  the inactive state of KaiC. Subsequently, KaiA is sequestered into a
  KaiA-KaiB-KaiC complex (ABC). {The total concentrations equal
    those used in the} {\em in vitro}{ experiments of}
  \cite{tomita05}{: ${\rm [C]_T = 0.58 \mu M; [A]_T = 1.75\mu M;
      [B]_T = 0.58\mu M}$, corresponding to $\rm [A]_T = [C]_T$ and
    $\rm [B]_T = 3 [C]_T$}{. For other parameter values, see Table
    1 of the} {\em Supporting Information}. \label{fig:fig3}
}
\end{figure}

\noindent {\bf KaiA + KaiB + KaiC}
Fig.~\ref{fig:fig3}A shows that our model produces {sustained}
oscillations in KaiC phosphorylation when all three Kai proteins are
present in the concentrations used in~\cite{kageyama06}. Both the
period and the amplitude of the oscillations agree well with those
observed in~\cite{nakajima05,kageyama06}.  Fig.~\ref{fig:fig3}B shows
the concentrations of complexes containing KaiA and KaiC ([AC]), KaiB
and KaiC ([BC]), and KaiA, KaiB, and KaiC ([ABC]), as a function of
time.  In the phosphorylation phase of the oscillations, KaiA binds to
KaiC and stimulates its phosphorylation. At the top
of the phosphorylation cycle, where KaiC hexamers flip from the active
to the inactive state, KaiA is released and KaiB binds to the inactive
KaiC hexamers. The binding of KaiB stabilizes the inactive form of
KaiC, preventing \phn\ by KaiA. One critical role of KaiB is thus to
allow the KaiC hexamers to enter the dephosphorylation phase of the
cycle.

Fig.~\ref{fig:fig3}B also shows that after [BC] has increased, [ABC]
increases. This is because ${\rm B_2}\widetilde{\rm C}_4 - {\rm
B_2}\widetilde{\rm C}_1$ can bind KaiA. This illustrates the second
function of KaiB: KaiB that is bound to KaiC also sequesters
KaiA. This leads to a form of the differential affinity mechanism at
the end of the dephosphorylation phase of the cycle: The KaiC hexamers
that are still in the inactive form---the laggards---will take away
KaiA from those hexamers that have already flipped from the inactive
to the active state---the front runners. This reduces the
phosphorylation rate of the front runners, allowing the laggards to
catch up.

{In our model, differential affinity acts at the
  bottom of the dephosphorylation phase of the cycle and throughout
  the phosphorylation phase.  From the perspective of synchronizing
  the oscillations of the different hexamers, the ideal would be an
  ever-decreasing affinity between KaiA and KaiC, even as a given
  hexamer passes through the same sequence of states again and
  again. Thermodynamics, however, dictates that the affinity of KaiA for
  KaiC must} {\em increase}{ somewhere in the cycle. In our model,
  this happens at the top of the inactive branch, where ${\rm
    B_2}\widetilde{\rm C}_6$ and ${\rm B_2}\widetilde{\rm C}_5$ do not
  bind KaiA, but ${\rm B_2}\widetilde{\rm C}_4$ does have a high
  affinity for KaiA.}

\begin{figure}[t]
\centering
\includegraphics[width=\columnwidth]{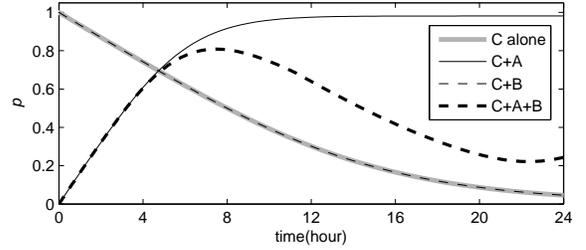}
\caption{
KaiC phosphorylation in the absence of KaiA and KaiB (C alone),
  in the presence of KaiA (C+A), in the presence of KaiB (C+B) and in
  the presence of both KaiA and KaiB (C+A+B). {For C+A, KaiC is
    initially fully unphosphorylated; for C alone and C+B, KaiC is initially
    fully phosphorylated (see also}
  \cite{kageyama06}). {Parameters as in Fig.} \ref{fig:fig3}.
\label{fig:fig4}
}
\end{figure}

\noindent {\bf KaiA + KaiC} Fig.~\ref{fig:fig4} {shows that, in the
presence of only KaiA, initially unphosphorylated KaiC reaches a
phosphorylation level of about 90-95\% after 6-8 hrs, in good
quantitative agreement with experiment}~\cite{kageyama06}. In our
model, KaiC is biased towards the active state, and KaiA binding
increases the stability of the active state even further.  This
explains the high steady-state phosphorylation level when only KaiA is
present.

\noindent {\bf (KaiB +) KaiC} Fig.~\ref{fig:fig4} {also shows that the
phosphorylation behavior of KaiC in the presence of KaiB is very
similar to that of KaiC alone, as observed}~\cite{xu03,kitayama03}. Our
model can explain this observation by assuming that the spontaneous
de\phn\ rate of the two KaiC conformations is the same and is
unaffected by KaiB binding, which only stabilizes the
inactive state with respect to the active one.

\subsection{Temperature compensation}

A striking feature of the {\em in vitro} oscillations of the Kai
system is that they are temperature compensated~\cite{nakajima05}.
Specifically, as the temperature is increased from $25^\circ$C to
$35^\circ$C, the period of the oscillations decreases by only
10\%. {In general, the oscillation period of a network depends upon
  the rates of all the reactions in the system. In principle, one
  could try to achieve temperature compensation by balancing the
  temperature dependences of all of these rates}
\cite{ruoff97}{{. We have adopted a different approach that is
    motivated by the fact that the (de)phosphorylation reactions are
    each individually temperature compensated}~\cite{tomita05}{:
    The phosphorylation time courses of KaiC alone and of KaiC with
    KaiA change little between $25^\circ$C and $35^\circ$C. Indeed,
    the key idea of our approach is to construct the model so that the
    oscillation period is determined by those rates that are known
    from experiment to be robust against temperature variations while
    leaving it insensitive to the other rates, which might vary with
    temperature.}

A natural idea is to demand that the rates that can vary with
temperature be much faster than the (de)phosphorylation rates, so that the
period is dominated by the latter, which are temperature
compensated. {This leads to:}\\
\noindent{\em 5. All (un)binding rates and the flip rates $f_6$ and
  $b_0$ are much faster than the (de)phosophorylation rates.}\\ Most
  conformational transitions are made at the top and bottom of the
  cycle; the period is thus less sensitive to flip rates other than
  $f_6$ and $b_0$.

Even when the (un)binding reactions between the Kai proteins are fast,
however, the period can still depend upon the {\em ratios} of their
rates---the dissociation constants---which will vary with
temperature. The period becomes independent of the dissociation
constants if all binding reactions go to completion. This occurs
when the dissociation constants are much smaller than typical
protein concentrations; in this limit, a change in the dissociation
constants will have no appreciable effect on the fraction of bound
proteins. We thus require:\\
\noindent{\em 6. The affinities among the Kai proteins are high.}\\
KaiA, the least abundant of the three proteins, will then
be almost entirely bound up in complexes with KaiB and KaiC, in
agreement with~\cite{kageyama06}. 
As long as
the {\em relative} magnitudes of the dissociation constants do not change with temperature, the
composition of these complexes will moreover be unaffected.
The phosphorylation rates, which depend on $[{\rm AC}_i]$,
are then robust to changes in temperature.  Another important consequence
of condition 6 is that a proportional increase in all of the protein
concentrations will have no effect on the oscillations, as has
been observed~\cite{kageyama06}.

\begin{figure}[t]
\centering
\includegraphics[width=\columnwidth]{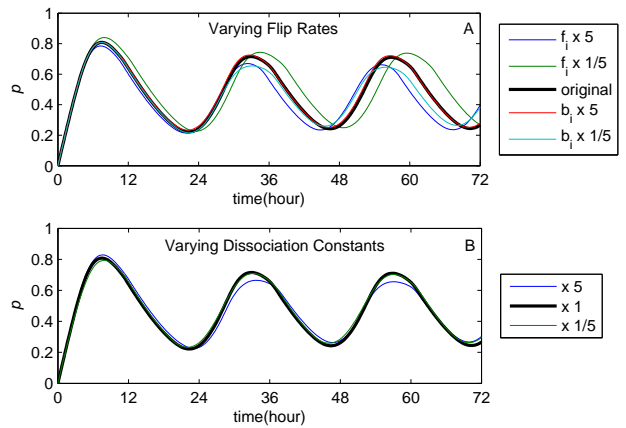}
\caption{
Temperature compensated oscillation
period. The period of KaiC phosphorylation changes by $10\%$ when
the forward ($f_i$) or backward ($b_i$) flip rates are changed
 by a factor 25 (A) and by less than 5\% when the
dissociation constants for all KaiA and KaiB binding reactions are
simultaneously changed by a factor $25$ (B). {Parameters as in Fig.} \ref{fig:fig3}.
\label{fig:fig5} 
}
\end{figure}

Since no data on the temperature dependence of the
dissociation constants and flip rates exists, we made the
following estimate:  We assumed that both the binding
energies and the energy barriers for the conformational transitions are
at most $50 k_{\rm B}T$.  If the temperature is changed
from $25^\circ$ to $35^\circ$, the dissociation constants and flip
rates can then change by about an order of magnitude. To test whether
our model is robust against such perturbations, we have varied
both dissociation constants and flip rates by a factor
five in each direction. Fig.~\ref{fig:fig5} shows that our model
withstands these trials:  The period varies by about $5-10\%$,
in very good agreement with experiment~\cite{nakajima05}.
This is strong evidence that conditions 5 and 6, together with
temperature-compensated (de)\phn\ rates, are sufficient for
temperature-compensated oscillations.

\subsection{KaiC dynamics as a function of the KaiA and KaiB
  concentration} 
  
\begin{figure}[t]
\centering
\includegraphics[width=\columnwidth]{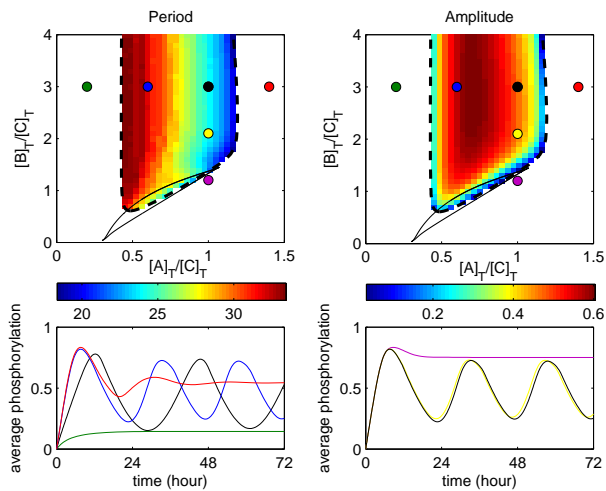}
\caption{
KaiC oscillations as a function of KaiA and KaiB
  concentration. (A) Period and (B) amplitude of oscillations in KaiC
  phosphorylation as a function of the concentration of KaiA and KaiB.
  The dashed curve shows the location of the supercritical Hopf
  bifurcation that gives birth to the oscillations, and the color
  scales give period in hours and amplitude of $p$ oscillation.
  {Note the small region of bistability (solid
    line; see also \hbox{\em Supporting Information}) at low ${\rm
      [A]_T}$ and ${\rm [B]_T}$. The remaining parameters are as in
    Fig.}~\ref{fig:fig3}. (C) KaiC oscillations as a function of KaiA
  concentration. Results are shown for ${\rm [B]_T=3[C]_T}$ and ${\rm
    [A]_T=0.2[C]_T}$ (green), ${\rm 0.6 [C]_T}$ (blue), ${\rm [C]_T}$
  (black) and ${\rm 1.4 [C]_T}$ (red). (D) KaiC oscillations as a
  function of KaiB concentration. Results are shown for ${\rm
    [A]_T=[C]_T}$ and ${\rm [B]_T=1.2[C]_T}$ (purple), $ {\rm 2.1
    [C]_T}$ (yellow) and ${\rm 3[C]_T}$ (black). \label{fig:fig6}
}
\end{figure}
  
Fig.~\ref{fig:fig6} shows the behavior of our model as a function of
the total KaiA and KaiB concentrations ${\rm [A]_T}$ and ${\rm
  [B]_T}$.  For ${\rm [A]_T} < 0.5 {\rm [C]_T}$, the system exhibits
no oscillations. At around ${\rm [A]_T = 0.5 [C]_T}$, the system
starts to oscillate {via a supercritical Hopf bifurcation}  with a period
of about 35 hrs {(see} {\em
  Supporting Information}{ for details on the bifurcation analysis)}. As the KaiA concentration is increased, the period
monotonically decreases. In contrast, the amplitude first increases to
reach a maximum at around ${\rm[A]_T = 0.85 [C]_T}$, then decreases
until oscillations disappear at around ${\rm [A]_T = 1.25 [C]_T}$. The
dynamics as a function of the KaiB concentration are markedly
different. Fig.~\ref{fig:fig6} shows that a minimum KaiB concentration
of about ${\rm [B]_T = [C]_T}$ is needed to sustain
oscillations.  Above that threshold, neither the period nor the
amplitude depend strongly on ${\rm [B]_T}$.

The different effects of varying ${\rm [A]_T}$ and ${\rm [B]_T}$ can
be understood from the different roles the two dimers play in our
model. KaiA stimulates the phosphorylation of KaiC.  If the total KaiA
concentration is very low, the phosphorylation rate will thus be so
slow that it is counterbalanced by the spontaneous dephosphorylation
rate.  If, on the other hand, the total KaiA concentration is very
high, the mechanism of differential affinity no longer functions,
because it relies on competition for a limited amount of KaiA. The
function of KaiB is to stabilize inactive KaiC and to sequester KaiA.
As long as enough KaiB is available to perform these functions, the
period and amplitude will not depend upon the KaiB concentration.

Interestingly, the very recent experiments of~\cite{kageyama06} give
strong support for our model. In particular, these experiments show
that when the KaiA and KaiB concentrations are reduced from their
standard values by a factor of four and three, respectively, all
oscillations cease in very good agreement with our results.  We
further predict that there is an upper bound on the KaiA
concentration, but not on the KaiB concentration, for oscillations to
exist.  Moreover, while the amplitude and the period of the
oscillations do not depend in our model on ${\rm [B]_T}$, they do
depend in a very specific manner on ${\rm [A]_T}$.  These dependences
on the KaiA and KaiB concentrations are direct consequences of the
basic roles of these proteins in our model. They thus represent some
of our most robust and important predictions.

\section{Discussion}
We have presented an allosteric model of KaiC phosphorylation that can
describe a wealth of experimental data on the Kai system. Its
foundation is the assumption that each KaiC hexamer can exist in two
distinct conformational states, an active one in which it tends to be
phosphorylated and an inactive one in which it tends to be
dephosphorylated. Because of the interplay between nucleotide binding,
which favors the active state, and phosphorylation, which favors the
inactive state, each individual hexamer will repetitively gain and
lose phosphate groups.  However, if macroscopic oscillations are to be
observed, the phosphorylation cycles of the individual hexamers must
be synchronized. We introduced a novel mechanism, called differential
affinity, which, in contrast to some previous
models~\cite{emberly06,mehra06}, allows for synchronisation even in
the absence of direct interactions between hexamers. The key idea is
that while all KaiC hexamers compete to bind KaiA, which stimulates
\phn, the laggards in the cycle are continuously being favored in the
competition. This mechanism is most effective when KaiB and KaiC bind
KaiA very strongly. It is also precisely in this limit that the
oscillation period becomes insensitive to changes in the Kai proteins'
affinities for each other.  Differential affinity and temperature
compensation are thus intimately connected.  The mechanism of driving
two-body reactions to saturation is{, however, more general; it could,
for instance, be used to make temporal programs of gene expression
robust against temperature variation} \cite{shenorr02}.

In {\em S. elongatus}, the concentration of KaiA dimers
is less than 10\% of that of KaiC hexamers~\cite{kitayama03}. Our
model predicts that in this regime the {\em in vitro} oscillations
of~\cite{nakajima05} disappear.  The very recent experiments
of~\cite{kageyama06} support this prediction: they unambiguously
demonstrate that {\em in vitro} the oscillations cease to exist if the
concentration of the KaiA dimers is less than 25\% of that of the KaiC
hexamers. Clearly, {\em in vivo} other processes are at work. It is known, for instance, that both the subcellular localization of the Kai
proteins~\cite{kitayama03} and KaiC's role as a transcriptional
repressor~\cite{nakahira04} affect
circadian rhythms, as do other clock proteins such as
SasA~\cite{kageyama03}.  It is tempting to speculate, however, that
these additional effects merely shift the phase boundaries of the
model presented here without changing its basic mechanism.  One could
imagine, for example, that a combination of KaiB localization to the
cell membrane and competitive binding by molecules like SasA could reduce the
number of sites available to sequester KaiA, thus allowing the
oscillator to function at lower KaiA concentrations.

Finally, our model makes a number of predictions that could be
verified experimentally. One clear prediction is that KaiC can exist in
two distinct conformational states.
Moreover, our model suggests that KaiC binds KaiA and KaiB very
strongly, with dissociation constants that depend upon the
conformational state and phosphorylation level of the KaiC hexamer.
But perhaps the strongest test of our model concerns the KaiC
oscillation dynamics as a function of the KaiA and KaiB concentrations
(see Fig.~\ref{fig:fig6}): We predict that the oscillations
will disappear when the KaiA concentration is increased, but not when
the KaiB concentration is increased.

We thank Daan Frenkel and Martin Howard for a critical reading of the
manuscript {and the Aspen Center for Physics for its hospitality to D.K.L. early in this project}. The work is supported by FOM/NWO.

\end{document}